\documentclass[conference]{IEEEtran}
\IEEEoverridecommandlockouts
\usepackage{cite}
\usepackage{amsmath,amssymb,amsfonts, amsthm}
\usepackage{algorithm}
\usepackage{algorithmicx}
\usepackage{algpseudocode}
\usepackage{graphicx}
\usepackage{textcomp}
\usepackage{xcolor}
\usepackage{booktabs}
\usepackage{multirow}
\usepackage{multicol}
\usepackage{diagbox}
\usepackage{url}
\usepackage{caption} 
\def\BibTeX{{\rm B\kern-.05em{\sc i\kern-.025em b}\kern-.08em
    T\kern-.1667em\lower.7ex\hbox{E}\kern-.125emX}}
\newtheorem{theorem}{Theorem}[section]

\begin{document}

\title{MCU-MixQ: A HW/SW Co-optimized Mixed-precision Neural Network Design Framework for MCUs\\

\author{
	\IEEEauthorblockN{Junfeng Gong$^{1,2}$, Cheng Liu $^{1,2}$\IEEEauthorrefmark{1}\thanks{\IEEEauthorrefmark{1} Corresponding author.}, Long Cheng$^{3}$, Huawei Li$^{1,2}$, Xiaowei Li$^{1,2}$}
	\IEEEauthorblockA{
		$^{1}$SKLP, Institute of Computing Technology, Chinese Academy of Sciences, Beijing, China
	}
	\IEEEauthorblockA{
		$^{2}$Dept. of Computer Science, University of Chinese Academy of Sciences, Beijing, China
	}
 \IEEEauthorblockA{
		$^{3}$School of Control and Computer Engineering, North China Electric Power University, Beijing, China
	}
 \vspace{-1em}

 \thanks{This work is supported by the National Key R\&D Program of China under Grant (2022YFB4500405), and the National Natural Science Foundation of China under Grant 62174162.}
}
}

\maketitle

\begin{abstract}
Mixed-precision neural network (MPNN) that utilizes just enough data width for the neural network processing is an effective approach to meet the stringent resources constraints including memory and computing of MCUs. Nevertheless, there is still a lack of sub-byte and mixed-precision SIMD operations in MCU-class ISA and the limited computing capability of MCUs remains underutilized, which further aggravates the computing bound encountered in neural network processing. As a result, the benefits of MPNNs cannot be fully unleashed. In this work, we propose to pack multiple low-bitwidth arithmetic operations within a single instruction multiple data (SIMD) instructions in typical MCUs, and then develop an efficient convolution operator by exploring both the data parallelism and computing parallelism in convolution along with the proposed SIMD packing. Finally, we further leverage Neural Architecture Search (NAS) to build a HW/SW co-designed MPNN design framework, namely MCU-MixQ. This framework can optimize both the MPNN quantization and MPNN implementation efficiency, striking an optimized balance between neural network performance and accuracy. According to our experiment results, MCU-MixQ achieves 2.1$\times$ and 1.4$\times$ speedup over CMix-NN and MCUNet respectively under the same resource constraints.

\end{abstract}



\section{Introduction}
The application of Artificial intelligence (AI) has become prevalent in typical Internet of Things (IoT) scenarios such as health monitoring, mechanical equipment fault diagnosis, and industrial automation. These applications commonly rely on microcontrollers (MCUs) known for their ultra-low power consumption and cost as the central processing units. Yet, AI especially deep learning models demands significant computational and memory resources, posing great challenges to its deployment on resource-limited MCUs. While embedding deep learning accelerators within MCUs is a conceivable strategy, it substantially escalates chip costs and energy consumption, thereby constraining its use in IoTs. Consequently, there is considerable interest in deploying lightweight deep learning models on MCUs to attain efficient inference and empower the intelligence of things\cite{banbury2021micronets, burrello2020dory, 10.1145/3517207.3526978}.

The two major challenges of deploying deep learning on MCUs are insufficient computational resources and limited memory capacity. For the memory issue, initiatives such as MCUNet\cite{lin2020mcunet, liberis2023pex} optimizes the memory scheduling during the inference from the perspective of the overall model topology. MCUNetV2 \cite{lin2021mcunetv2} further decomposes deep learning computations into smaller block computations to reduce the memory requirements. For the computation issue, neural network processing is known to be computing bound for MCUs and many efforts have been devoted to alleviate this challenge. Mixed-precision neural network (MPNN) quantization that can reduce both the computational and memory requirements of the models is a straightforward yet effective approach\cite{10509805}. However, MCUs generally lack support for sub-byte instructions\cite{9049084} and using primitive instructions for neural network processing directly would waste the limited computational resources in MCUs and aggravate the computing bound problem. 

Prior works \cite{lai2018cmsisnn, 9049084} investigated the use of SIMD instructions to accelerate low-bitwidth convolution operations, but they generally spread the low-bitwidth operations across the different SIMD lanes and the number of low-bitwidth operations packed in the SIMD is limited to the number of SIMD lanes. Essentially, they fail to make full use of the SIMD computing fabric because each SIMD lane is actually underutilized. In fact, the approach of packing low-bitwidth operations into high-bitwidth operations has also been explored for integer instructions of CPUs\cite{10.1109/ASP-DAC52403.2022.9712553} and primitive DSPs of FPGAs\cite{xilink8, xilink4, 10.1109/ASP-DAC52403.2022.9712553, 10035188}, but these approaches cannot be applied directly on SIMD instructions of MCUs. For instance, the shifting required by packing is almost free on FPGAs, but it is non-trivial for MCUs and inappropriate packing can even result in considerable performance penalty. 

To achieve efficient packing on MCUs, we propose to pack low-bitwidth operations with the granularity of the SIMD lanes such that each SIMD lane can be fully utilized. In addition, SIMD fabric usually enables different lane configurations, so we can also configure the SIMD lane sizes to suit the different bitwidth requirements of the convolution and achieve higher SIMD utilization. Furthermore, we rearrange the packing order of the low-bit operands to reduce the auxiliary instructions of packing and lower the packing overhead accordingly. 

As the bitwidth of the neural network models affects both the model accuracy and model implementation efficiency on MCUs, optimizing the model quantization and model implementation independently will lead to sub optimal results. In this work, we leverage NAS to take the model quantization and implementation efficiency of low-bitwidth operators into consideration at the same time and co-optimize the model accuracy and model performance. 
As mentioned, packing on MCUs requires additional shifting operations and the implementation efficiency of different bitwidth is not proportional to the bitwidth. To this end, we construct a performance model for neural network operators of low bitwidth such that the influence of different quantization on MCU implementation efficiency can be predicted immediately for the co-optimization. 
Finally, we have the SIMD-based convolution operator supporting various sub-byte operations added to TinyEngine\cite{lin2020mcunet} such that MPNNs can be deployed on top of TinyEngine efficiently and benefit the memory optimizations in TinyEngine.  

The main contributions of this work can be summarized as follows.
\begin{itemize}
    \item To enable efficient MPNN processing on MCUs, we propose an efficient low-bitwidth operation packing algorithm to make full into the SIMD fabric of MCUs. Specifically, we propose to have multiple low-bitwidth operations packed in the granularity of SIMD lanes and adapt the SIMD lane sizes to fit the convolution bitwidth requirements at the same time, achieving significantly higher packing efficiency compared to prior SIMD packing strategies. 
    
    \item Considering that the bitwidth of neural network models impacts both model accuracy and implementation efficiency, we leverage NAS to perform a packing-aware quantization for MCUs, thus co-optimizing model accuracy and performance concurrently. This framework is open sourced on GitHub \footnote{\url{https://anonymous.4open.science/r/MCU-MixQ-FCD7}}.

    \item With both the low-bitwidth packing and packing-aware NAS, we establish a HW/SW co-optimization framework, MCU-MixQ, for efficient neural network implementation on MCUs. According to our experiments on a set of neural network models, MCU-MixQ achieves $2.1\times$ and $1.5\times$ performance speedup over CMix-NN and MCUNet on average respectively under the same resource and accuracy constraints.

\end{itemize}


\section{Related Work}
To achieve efficient deep learning on MCUs with limited computing resources, prior work have proposed various optimization strategies from distinct angles. Early work mainly investigated mixed-precision quantization that explores smaller bitwidth for each different neural network layer to reduce both the compute and memory requirements. Some approaches concentrate on optimizing fundamental neural network operators by employing low-bitwidth operation packing and SIMD optimization, thereby increasing the implementation efficiency of the major neural network operators. Some of the research focused on HW/SW co-optimization, simultaneously considering operator implementation efficiency and model quantization. Additionally, there are also strategies \cite{lin2020mcunet} \cite{lin2021mcunetv2} targeting at memory optimization via scheduling and patching to accommodate larger deep learning models on MCUs with minimal performance loss. While the memory optimizations are generally orthogonal to the computing optimization, we mainly illustrate the computing optimization approaches in the rest of this section.

\subsection{Mixed-Precision Nerual Network Quantization}
Quantization is an established method for model compression, effectively reducing both computing and memory overhead. Currently, unified-precision quantization method has achieved remarkable success\cite{jin2022f8net}, even achieving lossless precision at 8-bit precision after fine-tuning\cite{pmlr-v129-li20a} or other novel techniques\cite{10376626}. Nevertheless, since model layers differ in their sensitivity to compression, conventional uniform precision quantization methods cannot realize optimal results. Several studies \cite{ZeroQ, HAWQ, pandey2023practical} have introduced various metrics to assess the sensitivity of different model layers, guiding the configuration of quantization schemes to strike a balance between accuracy and quantization bitwidth. It is a generic computing optimization approach for various computing engines including CPUs\cite{10.1109/ASP-DAC52403.2022.9712553}, FPGAs\cite{10.1109/ASP-DAC52403.2022.9712553, Chen_2021}, and GPUs\cite{kuzmin2022fp8, micikevicius2022fp8}, and can be particularly beneficial to MCUs with rather limited computing resources and energy budgets\cite{s21092984}.

\subsection{Network Operator Optimizations}
MPNNs that can reduce both the computing and memory requirements without compromising the model accuracy fit well with MCUs with limited hardware resources. Some recent works proposed to develop customized computing fabrics such as dot-product units and vector processing units to support convolution with lower bitwidth \cite{Bisdu} \cite{garofalo2020xpulpnn} \cite{garofalo2021ultralowpowercluster} in MPNNs, but there is still a lack of native low-bitwidth operator, particularly under 8 bit, in mainstream commodity MCUs. Implementing low-bitwidth deep learning models with primitive MCU arithmetic instructions directly leads to underutilization of the limited computing resources. To address this, CMSIS-NN \cite{lai2018cmsisnn} leverages SIMD instructions to optimize typical fixed point neural network operations like int8. CMix-NN \cite{9049084} investigates the use of SIMD instructions for 2 bit, 4 bit, and 8 bit convolution kernels for efficient MPNN inference. Hikconv \cite{10.1109/ASP-DAC52403.2022.9712553} presents a more general approach to pack arbitrary low-bitwidth convolution kernels on primitive integer arithmetic instructions in MCUs. The authors in \cite{TREC} explored the computing redundancy from inputs to improve the convolution performance.

\subsection{HW/SW Co-Optimization}
Despite of the improved computing and memory efficiency of MPNNs, the performance deployed on the target computing engines can vary substantially because of the implementation efficiency variations of the low-bitwidth neural network operations. To address the problem, hardware-aware quantization approaches have been explored to suit the different computing fabrics. While NAS \cite{NAS1000} \cite{liu2019darts} provides a unified design framework to search through the model design space for multi-objective optimization, it has been widely adopted to co-optimize the accuracy and performance of MPNNs in prior works \cite{cai2020rethinking}. TinyEngine \cite{lin2020mcunet} explored the model architecture along with the memory limitation to ensure effective neural network processing on MUCs.

In summary, recent studies have shown considerable promise for leveraging neural network redundancies through mixed-precision quantization and developing low-bitwidth convolution on current MCUs. However, dissociating quantization from operator optimization—targeting accuracy and performance separately—may result in suboptimal results. While several HW/SW co-optimization methods exist, there is still a lack of study on SIMD optimization for MPNNs, and prior efforts still fail to unleash the computing potential of MCUs. Innovative approaches that can both fully exploit MCU computing resources and concurrently compress model redundancies through quantization are highly demanded to perform neural network processing efficiently on MCUs.



\def\inlinecode#1{\texttt{#1}}
\section{Overall Design Framework}

\begin{figure*}[tb]
\centerline{\includegraphics[scale=0.4]{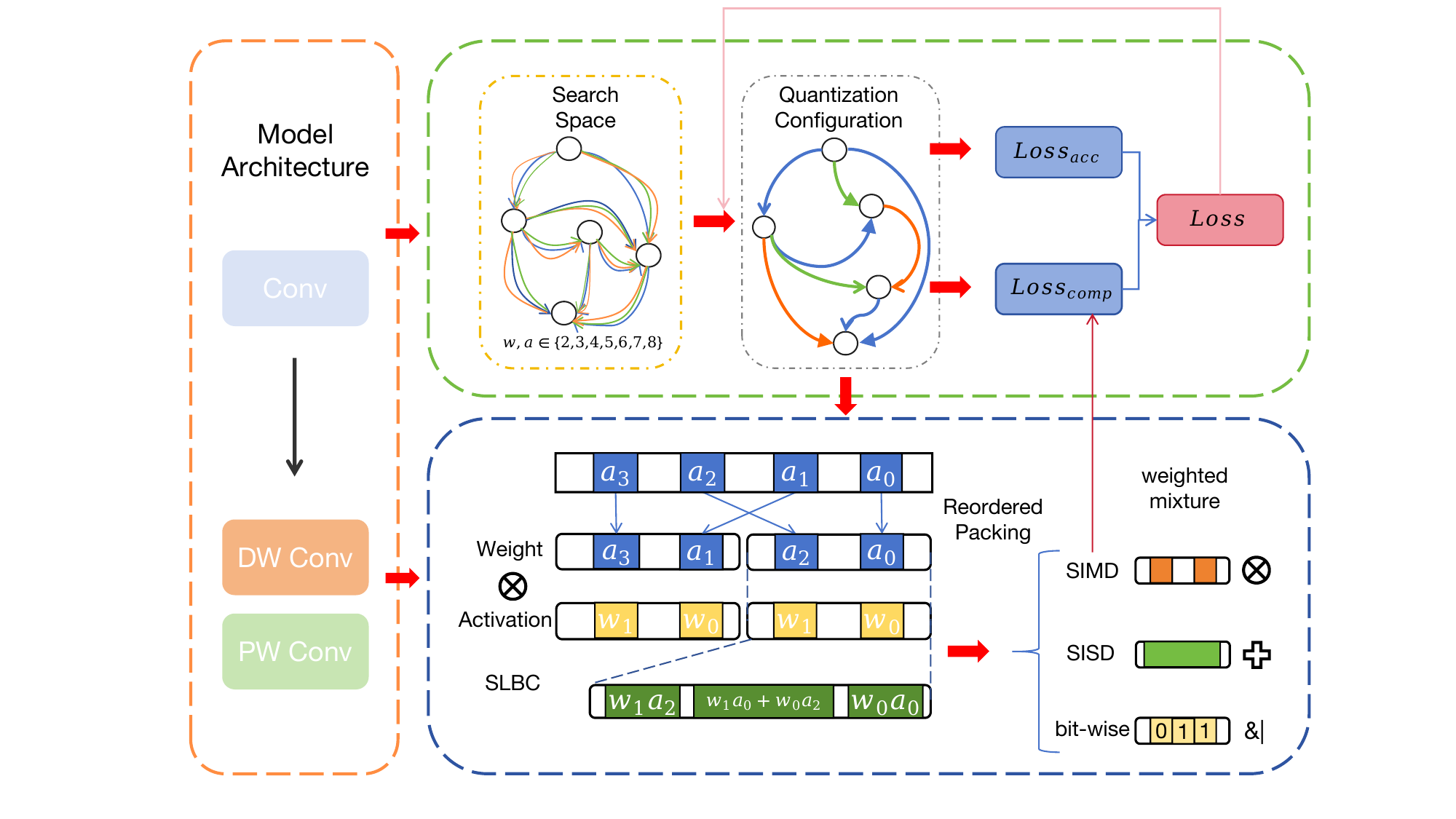}}
\caption{MCU-MixQ Overview, the proposed HW/SW co-optimization design framework for MPNN on MCUs.}
\label{fig:framework}
\end{figure*}

In this work, we present an MPNN design framework for MCUs, namely MCU-MixQ, as depicted in Fig. \ref{fig:framework}. This framework comprises SIMD-based low-bitwidth neural network operators and a hardware-aware quantization explorer based on NAS. The low-bitwidth neural network operator has multiple low-bitwidth operations packed into SIMD fabric, which makes full use of the computing resources in MCUs to mitigate the computing bottleneck. The hardware-aware quantization explorer is employed to reduce the data width of neural network models as much as possible, allowing for reduced computing and memory resource usage while retaining inference accuracy. Given that the implementation efficiency of low-bitwidth neural network operators also varies with the bitwidth configurations and significantly impacts network performance, the quantization explorer must be aware of the operator's implementation efficiency. This is achieved by incorporating a performance loss component alongside the standard accuracy loss component, as illustrated in Fig. \ref{fig:framework}.

\subsection{Low-bitwidth Network Operators}
To fully harness the computational resources, particularly SIMD, in MCUs for the acceleration of neural network processing, we seek to pack multiple low-bitwidth operations in a single SIMD fabric. Unlike prior SIMD packing \cite{8094835, 10.1145/3368826.3377912, 10.1109/ASP-DAC52403.2022.9712553} that fits each low-bitwidth operation to an independent SIMD lane, we propose to conduct the packing in each SIMD lane such that the packing is not limited to the minimum lane size i.e. 8-bit. Notably, we can adjust the SIMD lane sizes to the bitwidth requirements of the convolution and make best use of the SIMD fabric in MCUS. In addition, the packing typically requires shifting operations which also takes non-trivial overhead and affects the resulting performance of the network operators. To alleviate the packing overhead, we further propose a data reordering mechanism to reduce the number of auxiliary instructions required by the packing. The overall SIMD-based low-bitwidth convolution packing algorithm, namely SLBC, will produce an optimized sub-byte convolution operator, which will be utilized to sustain the execution of MPNNs. SLBC will be detailed in Section \ref{sec:SLBC}.

\subsection{Hardware-Aware MPNN Quantization}
Motivated by prior NAS-based quantization works \cite{cai2020rethinking, DQSS, DNAS, Edd, OSMPS}, we leverage a differentiable NAS to achieve hardware-aware quantization and co-optimize the model accuracy and performance. It starts with a pre-trained floating point model and sets the possible quantization data width as the initial design space of NAS. Then, it creates a quantization super-net to cover all the possible quantization configurations. Each layer of the target model to be quantized will be replaced with a mixed kernel composed of multiple weighted branches and each branch represents a specific quantization option. Given the quantization search space $Q=\{q_1,q_2,...,q_n\}$, the quantization super-net can be denoted as $f(Q)$, while a sub-net sampled from $Q$ is $f(q_i)$. The optimization goal is to search for a quantization sub-net $q^{*}$ to maximize the accuracy and minimize the latency while fulfilling the design constraints such as model sizes. 

With the super-net architecture, we can start the super-net training and have two loss components included to take both the model accuracy and model performance of different quantization setups into consideration in training as shown in Eq. \ref{eq:target1} and Eq. \ref{eq:target2}. 
Particularly, the performance loss component mainly characterizes the network performance when deployed on MCUs with the proposed SLBC packing approach. Since it is expensive to deploy the network with various quantization configurations on MCUs and extract the performance with realistic deployment, we have a simplified yet precise performance model for the NAS. The model is closely coupled with the SLBC packing and it will be illustrated in Section \ref{sec:SLBC} as well. After the quantization optimization, MCU-MixQ performs quantization aware training (QAT) on the selected mixed-precision model and the model will be deployed on MCUs, completing the entire workflow.

\begin{equation}
\label{eq:target2}
    Loss(\alpha_{w}, \alpha_{a})=\sum_{l=1}^{L}{C^{l}}
\end{equation}
\begin{equation}\label{eq:target1}
    Loss(\alpha_{w}, \alpha_{a})=Loss_{acc}(\alpha_{w}, \alpha_{a})+Loss_{comp}(\alpha_{w}, \alpha_{a})
\end{equation}

Finally, we deploy the obtained MPNN on MCUs with TinyEngine \cite{lin2020mcunet} which is an memory-efficient inference framework designed for MCUs. It provides all the major functionalities required to deploy a high-level model on MCUs. Particularly, it optimizes the memory usage of the model during the process of code generation and manages the memory scheduling to ensure on-demand parameter loading. These techniques prevent the out of memory issues during inference while minimizing its influence on the performance. While TinyEngine does not support sub-byte convolution operators, we have SLBC integrated to enable the deployment of MPNNs on TinyEngine. The revised TinyEngine in combination with the proposed quantization explorer constitutes the comprehensive HW/SW co-designed MPNN framework for MCUs known as MCU-MixQ.

\def\inlinecode#1{\texttt{#1}}

\section{Low-bitwidth Network Operator Optimizations} \label{sec:SLBC}
As a compute-intensive operator, convolution is heavily bounded by the computing capability of MCUs due to the lack of massively parallel computing fabrics. Therefore, optimizing the convolution, which is the major kernel of neural networks, is critical to the neural network performance on MCUs.

\subsection{SIMD-based Low-bitwidth Convolution}
Considering the mathematical equivalence of polynomial multiplication and convolution operation, for an $s_{b}$-bit sequence $s$ and a $k_{b}$-bit convolution kernel $k$, we can pack multiple low-bitwidth elements of $s$ and $k$ into one wider hardware unit $R_1$ and $R_2$ which can be represented with the following polynomial forms.
\begin{align}
    R_1&=\sum_{i=0}^{N_{s} - 1}{s[i]\cdot 2^{iS_{b}}} \\
    R_2&=\sum_{j=0}^{N_{k} - 1}{k[j]\cdot 2^{jS_{b}}}
\end{align}
With the packing, the product $P$ of a high-precision multiplier can be simplified to Equation \ref{eq:product} according to the rule of polynomial multiplication.
\begin{align}
    P&=R_1\times R_2 \notag \\
    &=(\sum_{i=0}^{N_{s} - 1}{s[i]\cdot 2^{iS_{b}}})\cdot(\sum_{j=0}^{N_{k} - 1}{k[j]\cdot 2^{jS_{b}}}) \label{eq:product} \\
    &= \sum_{k=0}^{N_{s}+N_{k}-2}{(\sum_{i+j=k}{s[i]\cdot k[j]\cdot 2^{kS_{b}}})} \notag
\end{align}
According to the definition of convolution, the application of the $N_{k}$-kernel $k$ to a $N_{s}$-element sequence $s$ also yields $N_{s}+N_{k}-1$ elements and the $n$th element of convolution sequence $y$ can be represented as:
\begin{align}
    y[n]&=\sum_{m=0}^{N_{k}-1}{s[n-m]\cdot k[m]} \label{eq:convolution_sequence} \\
    &=\sum_{i+j=n}{s[i]\cdot k[j]} \notag
\end{align}
According to Eq. \ref{eq:product} and Eq. \ref{eq:convolution_sequence}, it is evident that the multiplication product $P$ is composed of convolution sequence $y$, each of which has been left-shifted by the corresponding number of bits with Eq. \ref{eq:representation}. As a result, each element of convolution sequence can be segmented from $P$ through bit operations. By utilizing a single multiplication instruction along with multiple bit-wise instructions for packing and segmentation, the overhead of the convolution can be significantly reduced compared to naive implementation.
\begin{align}
    P&=\sum_{k=0}^{N_{s}+N_{k}-2}{y[k]\cdot 2^{k_{b}}} \label{eq:representation}
\end{align}

\begin{figure}[tb]
\centerline{\includegraphics[scale=0.25]{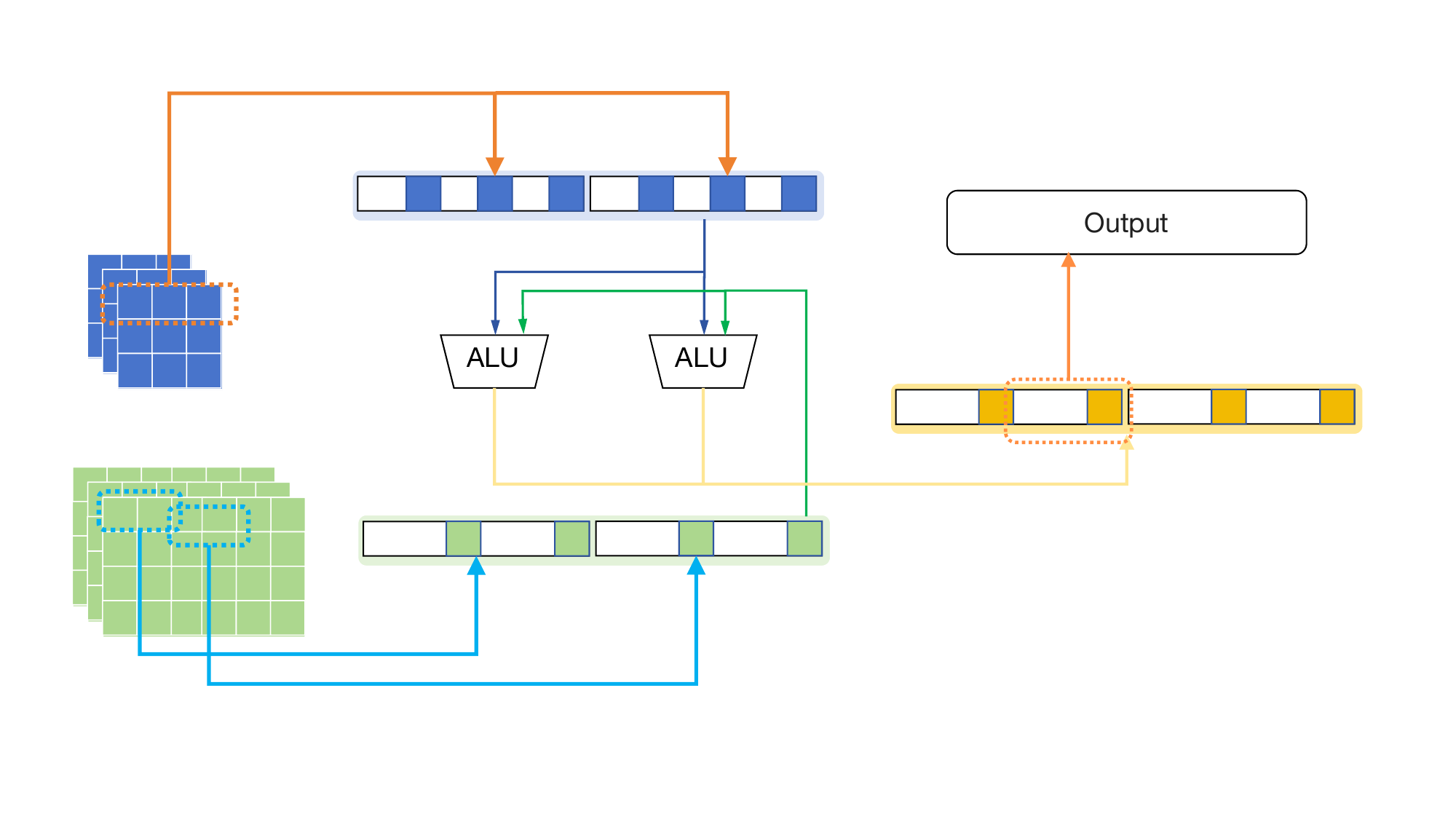}}
\caption{SIMD Low-bitwidth Convolution (SLBC) Overview}
\label{fig:SLDC_overview}
\end{figure}

The above analysis assumes that the computation unit is long enough to accommodate all elements from the sequence and kernel. In practice, most commodity MCUs like Cortex-M incorporate SIMD instructions to enhance parallel computing capabilities. Thus, we investigate the use of the SIMD instructions for efficient low-bitwidth convolution and propose SLBC, a \textbf{S}IMD \textbf{l}ow-\textbf{b}itwidth \textbf{c}onvolution optimized for MCUs. 
The detailed execution flow of SLBC is presented in Algorithm \ref{alg:SLBC}. With SLBC, multiple multiply and add operations in a convolution operator can be substituted with a single SIMD multiplication instruction and bit-wise operations.
SLBC mainly consists of three processing stages including packing, SIMD multiplication, and SIMD segmentation as illustrated in Fig. \ref{fig:SLDC_overview}. 

\begin{algorithm}
\caption{Naïve SLBC}
\label{alg:SLBC}
\begin{algorithmic}
\For{$i=0$ to $K$ by $1$} \Comment{kernel packing}
    \State $B \mathrel{\vert}= k[i] \ll (i \times S)$
\EndFor
\State
\State $mask=(1 \ll S) - 1$
\State $vmask=vdupq\_n(mask)$ \Comment{vector mask}
\State
\For{$i=0$ to $N_f$ by $N\times2$} \Comment{elements packing}
    \For{$i=0$ to $G$ by $1$}
        \For{$k=0$ to $N$ by $1$}
            \State $A[j] \mathrel{\vert}= A[i+j\times N+k] \ll (k\times S)$
        \EndFor
    \EndFor
    \State
    \State $VA=vld(A)$
    \State $VP=vmul(VA, VB)$ \Comment{SIMD multiplication}
    \State
    \For{$j=0$ to $N+K-1$ by $1$} \Comment{extract}
        \State $shift\_vp = vshr(VP, vmask)$
        \State $vresult=vand(shift\_vp, vmask)$
        \State
        \State $scalar0=vget(vresult, 0)$
        \State $scalar1=vget(vresult, 1)$
        \State
        \State $output[i+0\times N+j] \mathrel{+}=scalar0$
        \State $output[i+1\times N+j] \mathrel{+}=scalar1$
    \EndFor
\EndFor
\end{algorithmic}
\end{algorithm}

In packing stage, multiple convolution elements can be packed into a wider SIMD register file with SIMD shift instruction and vector mask operation. Suppose each SIMD register has $N_{l}$ lanes and each lane is able to pack $N_{s}$ sequence elements and $N_{k}$ kernel elements. When $N_{k}$ is equal to the kernel size $k$, the entire kernel can be packed into a single SIMD lane. According to the packing strategy defined previously, the packed SIMD register $vs$ and $vk$ can be denoted as Eq. \ref{eq:vs} and Eq. \ref{eq:vk} respectively.
\begin{align}
    vs&=\sum_{l=0}^{N_{l}-1}{(2^{lL_{b}}\cdot (\sum_{i=0}^{N_{s}-1}{s[lN_{s}+i]\cdot 2^{iS_{b}}}))} \label{eq:vs} \\
    vk&=\sum_{l=0}^{N_{l}-1}{(2^{lL_{b}}\cdot (\sum_{i=0}^{N_{k}-1}{k[i]\cdot 2^{iS_{b}}}))} \label{eq:vk}
\end{align}

In SIMD multiplication stage, the packed data $vs$ and $vk$ multiply with an SIMD instruction and the product is presented in Eq. \ref{eq:SLBC}. Note that $\otimes$ denotes the SIMD multiplication and $N_l$ denotes the total number of SIMD lanes. After the SIMD multiplication, the convolution sequence is already stored in the output vector, which means that we can replace more \inlinecode{ADD} and \inlinecode{MUL} i.e. single instruction single data (SISD) instructions with one SIMD instruction. 
\begin{align}
    vp&=vs\otimes vk \label{eq:SLBC} \\
    &=\sum_{l=0}^{N_{l}-1}{(2^{lL_{b}}\cdot (\sum_{i=0}^{N_{s}-1}{s[lN_{s}+i]\cdot 2^{iG_{b}}}))} \notag \\
    &\otimes \sum_{l=0}^{N_{l}-1}{(2^{lL_{b}}\cdot (\sum_{i=0}^{N_{k}-1}{k[i]\cdot 2^{iG_{b}}}))} \notag \\
    &= \sum_{l=0}^{N_{l}-1}{(2^{lL_{b}}\cdot \sum_{k=0}^{N_{s}+N_{k}-2}{(\sum_{i+j=k}{(vs[lN_{s}+i]\cdot vk[j]\cdot 2^{kG_{b}}}))})} \notag
\end{align}

In segmentation stage, we notice that SLBC can be viewed as multiple parallel packing tasks. As shown in Eq. \ref{eq:SLBC_segmentation}, the last element in a lane will be combined with the first element of the next lane to form an element of the convolution sequence, while the other data in each SIMD lane also become elements of the convolution sequence. Note that $vp$ represents a 2-D array while the first dimension represents the lane index and the second index represents the element position in each lane. Finally, SIMD bit-wise operations is utilized to extract the convolution sequence from the output vector.

\begin{equation}
y[i]=\left\{
\begin{aligned}
& vp[l][N_{s}+N_{k}-2]+vp[l+1][0], i\ne0,i\ne N-1,k=0 \\
& vp[l][k], others
\end{aligned}
\right.
\label{eq:SLBC_segmentation}  
\end{equation}
subject to \par
$$
\left\{
\begin{aligned}
l &= i / (N_{s}+N_{k}-2) \\
k &= i \% (N_{s}+N_{k}-2) \\
N &= lN_{s}+N_{k}-1
\end{aligned}
\right.
$$

\subsection{Enhance Locality Through Reordering}
Despite the packing efficiency, SLBC requires extra bit operations such as \inlinecode{LSR} to extract convolution elements from the output vector and the overhead of these bit operations is non-trivial. Inspired by ULPPACK \cite{ULPPACK} that utilizes local accumulation to combine multiple bit operations together and reduces segmentation overhead substantially, we propose a new reordering algorithm for SLBC to improve the register reuse during packing, as shown in Theorem \ref{the:reordered_packing}. 
 
\begin{theorem}
\label{the:reordered_packing}
For SIMD registers with $L$ lanes, each lane can accommodate $N$ low-bitwidth elements, a group of $N\times L^{2}$ elements will be reordered and packed within $L$ SIMD registers. For the $y$th lane of the $x$th SIMD register, it will be packed into the $y$th position of the $x$th one.
\end{theorem}

\begin{figure}[tb]
\centerline{\includegraphics[scale=0.28]{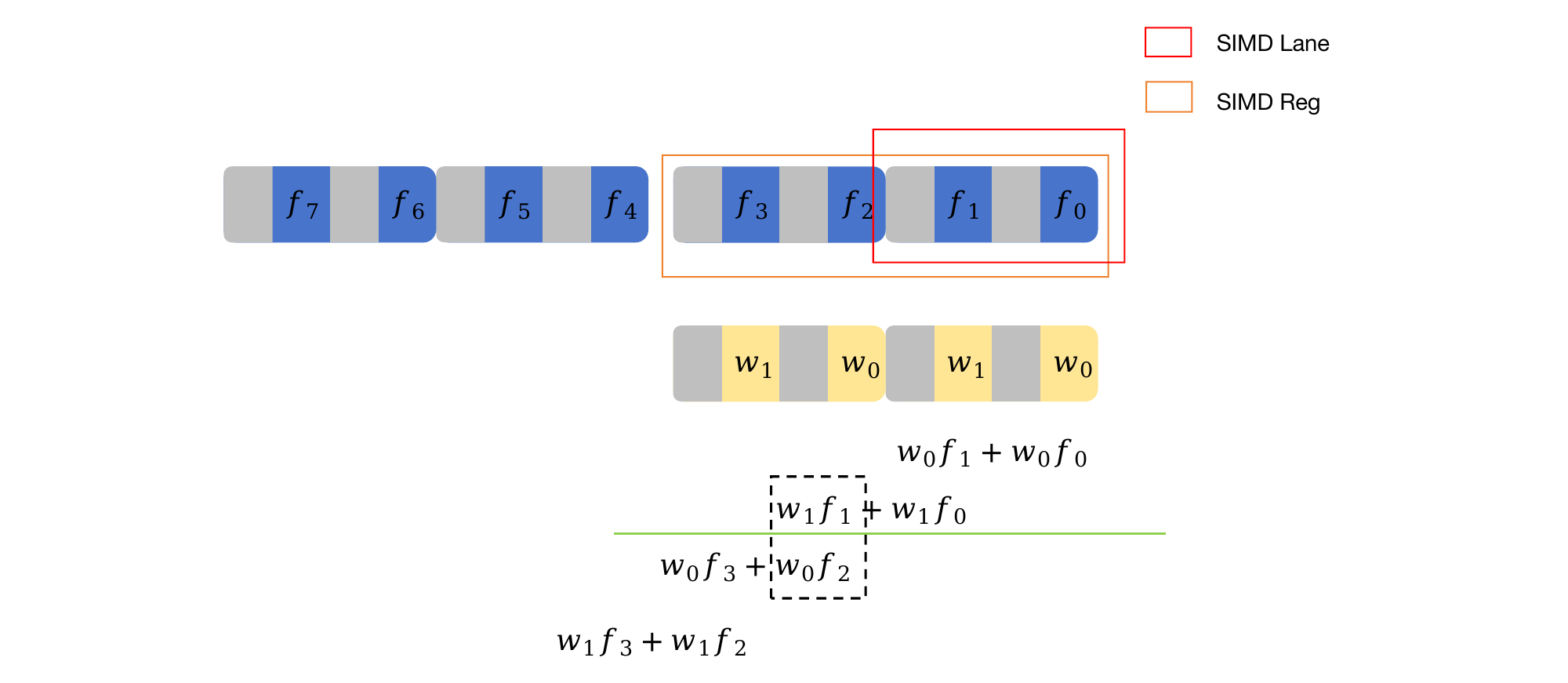}}
\caption{Output overlapping in naïve SLBC}
\label{fig:naive_SLDC}
\end{figure}

\begin{figure}[tb]
\centerline{\includegraphics[scale=0.28]{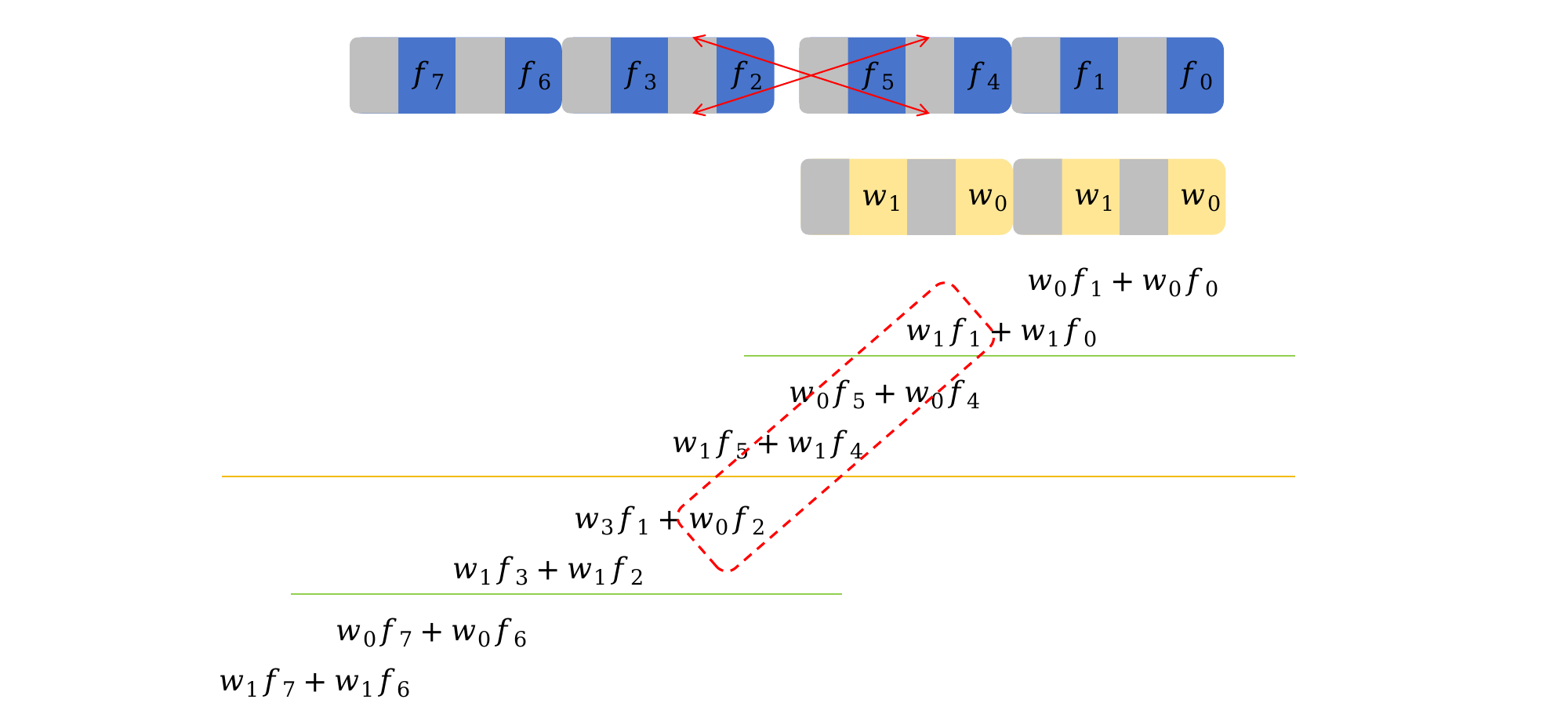}}
\caption{Output Overlapping in Reordered SLDC}
\label{fig:reordered_SLDC}
\end{figure}

To illustrate the reordering algorithm, we have two simplified packing examples presented in Fig. \ref{fig:naive_SLDC} and Fig. \ref{fig:reordered_SLDC}. Suppose each SIMD register has 2 lanes and each lane can pack 2 elements. Assume that the kernel size is also 2 so that the entire kernel can be fully packed into one SIMD lane. Fig. \ref{fig:naive_SLDC} shows the processing of the naive packing proposed in Algorithm \ref{alg:SLBC_reordered}. Since the entire kernel can be packed into one lane and convolution sequence needs to be segmented, two different segmentations of sequence will be packed into one SIMD register and meanwhile the entire kernel will be packed into each lane of the same SIMD register to perform an optimized convolution through SIMD multiplication. But according to the details of SLBC, the overlapping part are distributed in adjacent lanes within the same SIMD register. According to the principle of SIMD, the overlapping part can not be utilized through shift operation. As a result, the overlapping part needs to be segmented separately from adjacent lanes, thus leading to unnecessary overhead in bit-wise operation instructions.

In order to fully utilize the overlapping portions and merge multiple segmentation operations together, the arrangement order of elements has been modified so that the overlapping portions appear in adjacent SIMD registers rather than between adjacent lanes within the same one. The specific method of reordered packing SLBC is shown in Fig. \ref{fig:reordered_SLDC}. It can be observed that since the order of data rearrangement has been changed, there exists overlap between the result of adjacent iterations. For SIMD registers with $L$ lanes that store the product results, which represent the convolution sequence after packing, the element contained in the boundary lane cannot form one complete element of the convolution independently. Instead, it needs to be added to the element held in the first lane in the adjacent SIMD register to become an real element. In other words, the elements located in boundary position require an additional segmentation operation. Fig. \ref{fig:reordered_SLDC} illustrates the packing positions of these two elements in SIMD registers. However, After rearranging the packing order of elements, the boundary elements to jointly form one complete convolution element are located in corresponding lanes of adjacent SIMD registers. Therefore, these two SIMD registers can be accumulated after performing parallel shifting operations, which eliminates the need for additional splitting overhead. For the configuration discussed above, $L$ segmentation operations will be eliminated for every $N\times L\times L$ elements, thus reducing segmentation overhead to $\frac{1}{N\times L}$ of the original count. 

\par

\begin{algorithm} 
	\caption{SLBC with reordered packing}
	\label{alg:SLBC_reordered} 
	\begin{algorithmic}
            \For{$i=0$ to $N_f$ by $N\times G\times G$}
                \State $local=vdup(0)$ \Comment{local accumulator}
                \For{$j=0$ to $G$ by $1$} \Comment{reordered packing}
                    \For{$k=0$ to $K$ by $1$}
                        \For{$l=0$ to $N$ by $1$}
                            \State $A[k]=f[i+j\times N+4\times k+l] \ll (l\times S)$
                        \EndFor
                    \EndFor
                    \State
                    \State $VA=vld(A)$
                    \State $VP=vmul(VA, VB)$
                    \State $vst(local, VP)$
                    \State $local=vadd(local, VP)$
                    \State
                    \For{$k=0$ to $N$ by $1$}
                        \State $shift_vp=vshr(local, k\times S)$
                        \State $vresult=vand(shift_vp, vmask)$
                        \State $scalar0=vget(vresult, 0)$
                        \State $scalar1=vget(vresult, 1)$
                        \State $output[i+j\times N+k]\mathrel{+}=scalar0$
                        \State $output[i+j\times N+G\times N+k]\mathrel{+}=scalar1$
                        
                    \EndFor
                    \State
                    \State $local=vshr(local, N\times S)$
                \EndFor

                \For{$j=0$ to $K-1$ by $1$}
                    \State $scalar0=vget(local, 0)$
                    \State $scalar1=vget(local, 1)$
                    \State $output[i+G\times N+j]\mathrel{+}=scalar0$
                    \State $output[i+2\times G\times N+j]\mathrel{+}=scalar1$
                \EndFor
            \EndFor
	\end{algorithmic} 
\end{algorithm}

The complete algorithm for reordered SLBC is illustrated in Algorithm.\ref{alg:SLBC_reordered}. During packing stage, $N\times L\times L$ elements will be considered as a group to be packed into $N$ SIMD registers. After the completion of the multiplication operation, the convolution elements squeezed in the register will not be segmented immediately. Instead, it will be right-shifted and added to the local accumulation after each round of multiplication. After the group of multiplications is completed, the actual elements will be segmented from the local accumulation. In the subsequent experimental phase, we conducted an ablation study on reordered SLBC and SLBC within an end-to-end framework to validate the effectiveness of the improved method in reducing segmentation overhead.

\subsection{Adaptive SIMD Packing}
SLBC has low-bitwidth operations packed into each SIMD lane independently, but the packing efficiency depends on both the SIMD lane size and the operation bitwidth to be packed. Since SIMD usually allows different lane sizes, we can adjust the SIMD lane size to fit the convolution bitwidth for higher SIMD utilization. For each convolution of an MPNN, we adaptively decide the optimized packing and SIMD lane sizes at compilation time to ensures optimized MPNN performance.

\subsection{Packing Performance Prediction}
 
The proposed HW/SW co-design framework MUC-MixQ requires a large number of performance evaluation of MPNNs with different quantization setups which can be too expensive for evaluation with realistic deployment, so we further build a performance model for this purpose. As mentioned, the low-bitwidth convolution implemented with SLBC includes both SIMD \inlinecode{MUL} instruction and bitwise operations. Considering the varied execution time of the different types of instructions, we use SISD instructions as the calibration metric and align SIMD \inlinecode{MUL} instruction and SISD bit operations with it. Specifically, as shown in Eq. \ref{eq:total_complexity}, the complexity of SISD instructions $C_{SISD}$ is roughly proportional to the number of SISD accumulation and multiplication operations where $\alpha$ and $\beta$ refers to the proportion coefficients and they can be obtained with experiments.


\begin{equation}
    C=C_{SISD}+\alpha C_{SIMD}+\beta C_{bit}
\label{eq:total_complexity}
\end{equation}


\section{Experiment}
To showcase the outstanding performance of MCU-MixQ on MCUs, we conducted experiments on two datasets: Visual Wake Word (VWW) and CIFAR-10. VWW is a vision-oriented dataset specifically designed to determine the presence or absence of a person in an image. CIFAR-10 is a widely adopted benchmark for image classification tasks. For the hardware platform, we selected ARM Cortex-M7 microcontroller STM32F746, which is equipped with 320kB of SRAM and 1MB of Flash memory. All the latency measurement is obtained at a clock frequency of 216MHz.

\subsection{End-to-End Performance Evaluation}
We have the neural network benchmark implemented on the target hardware platform with CMix-NN\cite{9049084}, WPC\&DDD\cite{9881566}, TinyEngine\cite{lin2020mcunet} and the proposed MCU-MixQ respectively. Note that CMix-NN and WPC\&DDD only supports three different bitwidth setups i.e. 2bit, 4bit, and 8bit, TinyEngine only supports 8bit while MCU-MixQ supports all the bitwidth between 2bit and 8bit. Given the same model accuracy constraint, we compared the end-to-end performance of the resulting neural network models and the comparison is summarized in Table \ref{tab:end_to_end}. It can be observed that MCU-MixQ achieves the best performance and outperforms all the other solutions. This can be attributed to multi-folded reasons including the more efficient low-bitwidth convolution optimization and more flexible quantization, as well as the HW/SW co-optimization. They will be analyzed in detail in the rest of the experiments. Moreover, MCU-MixQ takes advantage of the memory optimization provided by TinyEngine, so the peak memory usage is also reduced. On the other hand, we notice that CMix-NN and WPC\&DDD with more flexible quantization setups show even lower performance than TinyEngine with fixed int8 quantization. This is mainly attributed to other optimization techniques introduced by TinyEngine, model-adaptive memory scheduling and 
computation kernel specialization for example.

\begin{table*}[tb]
    \centering
    \caption{End-to-end performance comparison with previous frameworks}
    \begin{tabular}{c|c|c|c|c|c|c|c}
        Backbone & method & Quantization & Peak Memory & Flash Memory & Clocks & Latency & Accuracy\\
        \hline
        \multirow{4}{*}{VGG-Tiny} & CMix-NN\cite{9049084} & Mixed(2,4,8) & 146.33KB & 146.33KB & 5680854 & 26.3ms & 71.4\% \\
        \multirow{4}{*}{} & WPC\& DDD\cite{9881566} & Mixed(2,4,8) & 228.59KB & 146.33KB & 5140887 & 23.8ms &  70.2\% \\
        \multirow{4}{*}{} & TinyEngine\cite{lin2020mcunet} & 8-bit & 51.93KB & 584.65KB & 3715233 & 17.2ms & 78.1\% \\
        \multirow{4}{*}{} & MCU-MixQ & Mixed(2-8) & 49.92KB & 591.12KB & 2721615 & 12.6ms & 78.3\% \\
        \hline
        \multirow{4}{*}{MobileNet-Tiny} & CMix-NN\cite{9049084} & Mixed(2,4,8) & 184.20KB & 184.20KB & 9136773 & 42.3ms & 76.2\% \\
        \multirow{4}{*}{} & WPC\& DDD\cite{9881566} & Mixed(2,4,8) & 272.38KB & 184.20KB & 7819221 & 36.2ms & 74.3\% \\
        \multirow{4}{*}{} & TinyEngine\cite{lin2020mcunet} & 8-bit & 68.37KB & 704.15KB & 5983279 & 27.7ms & 80.1\% \\
        \multirow{4}{*}{} & MCU-MixQ & Mixed(2-8) & 62.14KB & 687.36KB & 4600823 & 19.5ms & 79.7\% \\
    \end{tabular}
    \label{tab:end_to_end}
\end{table*}

\subsection{SLBC Efficiency Evaluation}
\begin{figure}[hbtp]
\centerline{\includegraphics[scale=0.26]{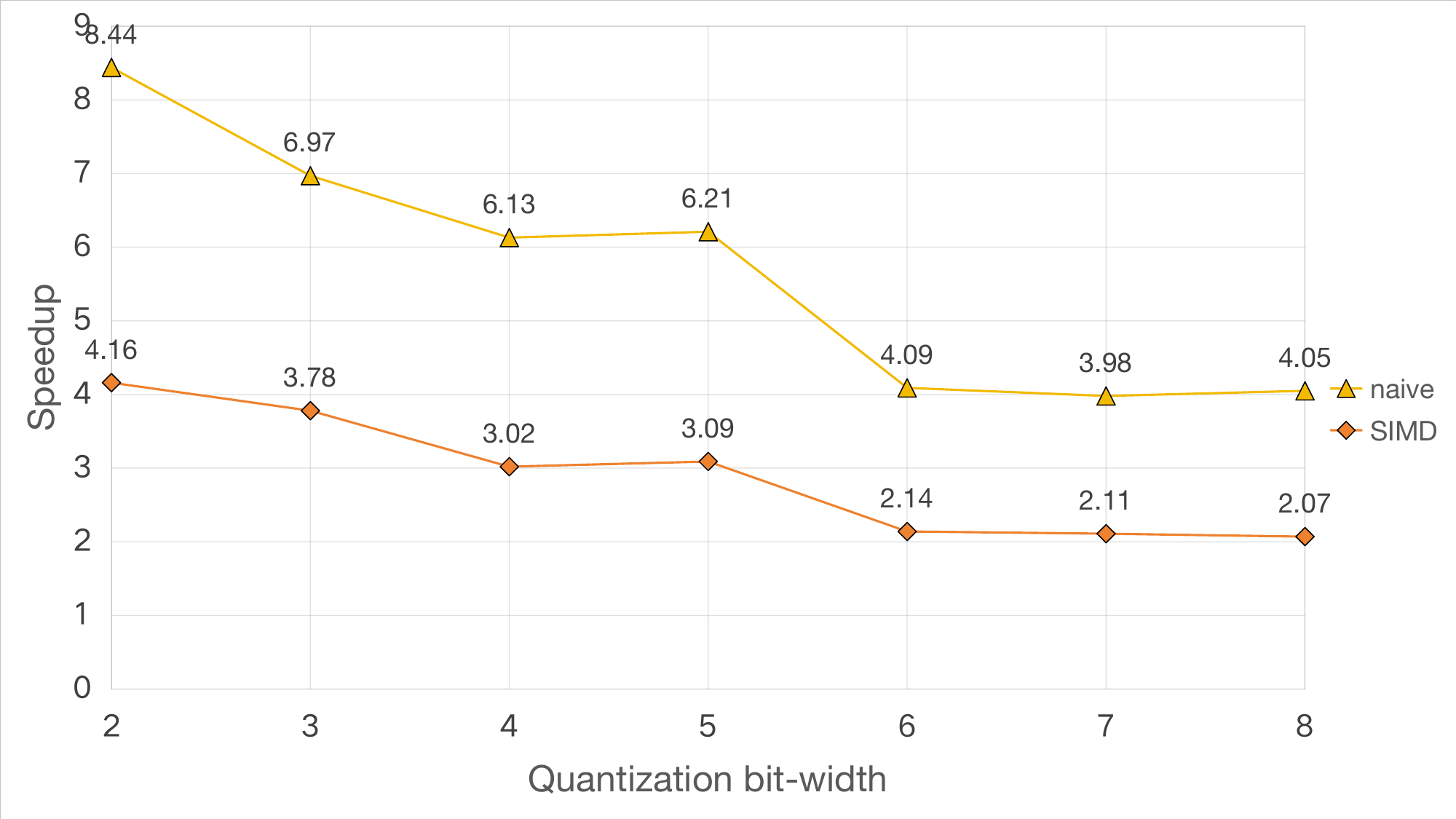}}
\caption{Speedup over naive and SIMD convolution}
\label{fig:comp}
\end{figure}

First of all, we compare SLBC with other convolution kernels. In order to showcase its efficacy on low-bit convolutions, we compare SLBC with naive convolution, SIMD convolution and CMix-NN. SIMD convolution uses SIMD instructions to accelerate convolution without other optimization method. Due to the lack of support for sub-byte in naive and SIMD convolution, the latency of the convolution operator does not change when executing with different bitwidths under 8 bits. Fig \ref{fig:comp} illustrates the speedups of SLBC under different bitwidths over the two methods. According to the experimental results, SLBC achieves an average speedup of $4\times$ and $2\times$ over naive and SIMD convolution seperately. \par

\begin{figure}[hbtp]
\centerline{\includegraphics[scale=0.25]{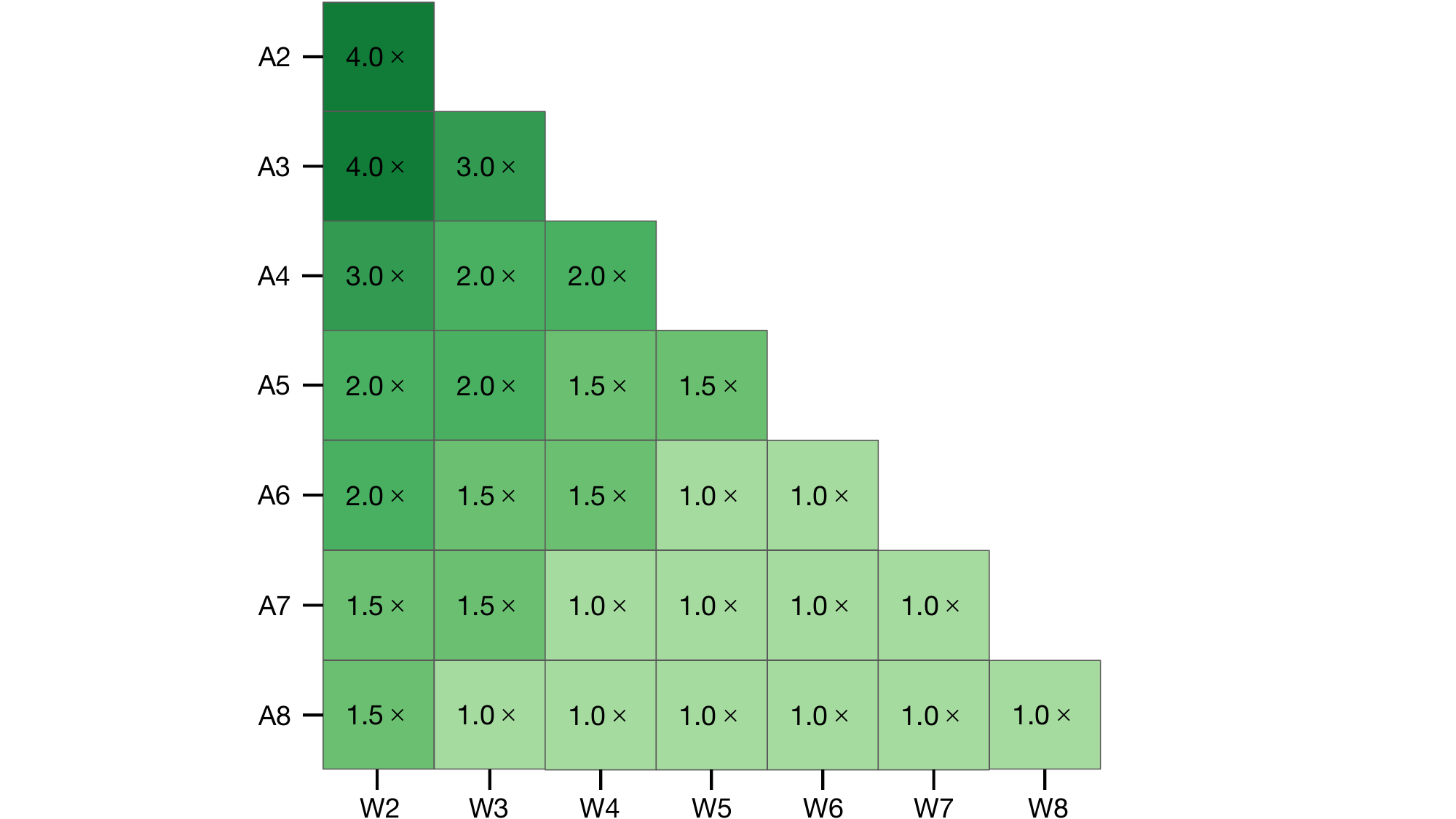}}
\caption{Speedups over CMix-NN}
\vspace{-1.5em}
\label{fig:conv_kernel_comp}
\end{figure}

CMix-NN is a flexible mixed-precision inference library, which supports any combination of 2, 4, 8 bitwidth. It compresses low-bitwidth data for storage and simultaneously constructs vector instructions using masks in convolution. In order to demonstrate the superiority of SLBC over CMix-NN in terms of hardware resource utilization, we compared the theoretical throughput of the two methods. More specifically, Fig \ref{fig:conv_kernel_comp} presents the acceleration ratios for different bitwidth combinations, which represents the equivalent ratio of operations performed by the one SIMD instruction. According to Fig \ref{fig:conv_kernel_comp}, SLBC can achieve up to $1.5\times$ speedup over CMix-NN in most quantization combination.

\begin{figure}[hbtp]
\centerline{\includegraphics[scale=0.25]{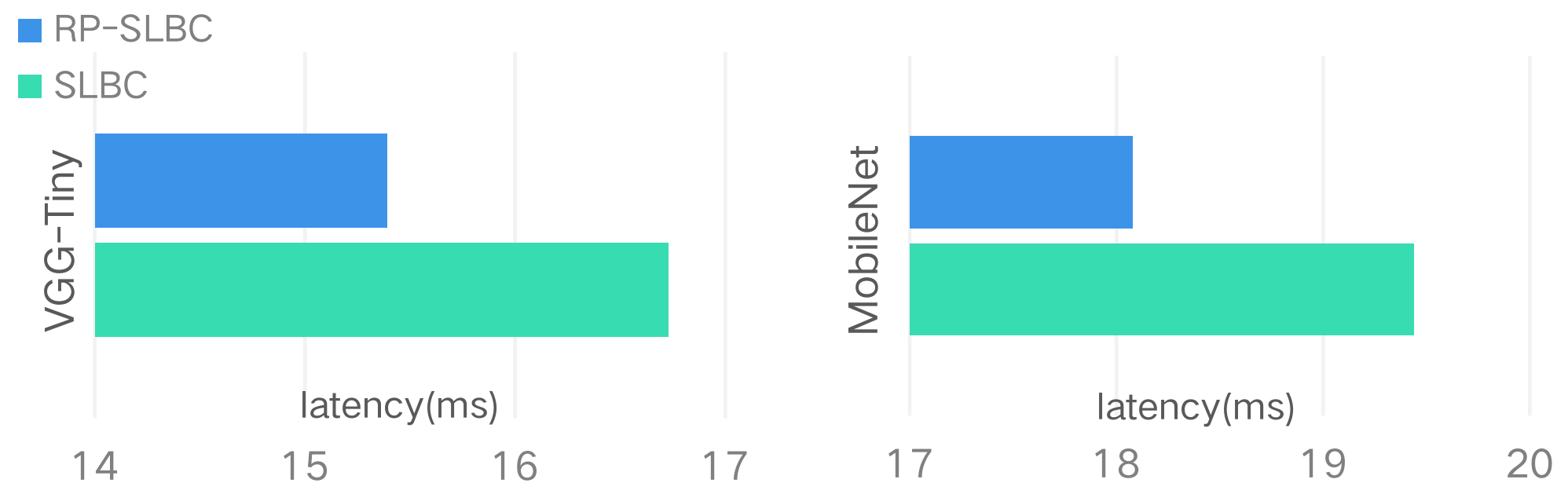}}
\caption{Latency comparison between SLBC and RP-SLBC}
\vspace{-1em}
\label{fig:ablation}
\end{figure}

To demonstrate the effectiveness of our reordered packing SLBC(RP-SLBC), we conducted ablation experiments on the new method. More specifically,
we integrate SLBC and RP-SLBC into our end-to-end deep learning framework respectively. For each convolution kernel, we calculate their complexity according to Algorithm \ref{alg:SLBC} and Algorithm \ref{alg:SLBC_reordered} separately, and evaluate the model on MCU platform. Based on the results in Figure. \ref{fig:ablation} and theory analysis in Algorithm \ref{alg:SLBC}, RP-SLBC can reach up to nearly $1.1\times$ through reordered packing.

\subsection{HW/SW Co-optimization Evaluation}
\begin{figure}[htbp]
\centerline{\includegraphics[scale=0.25]{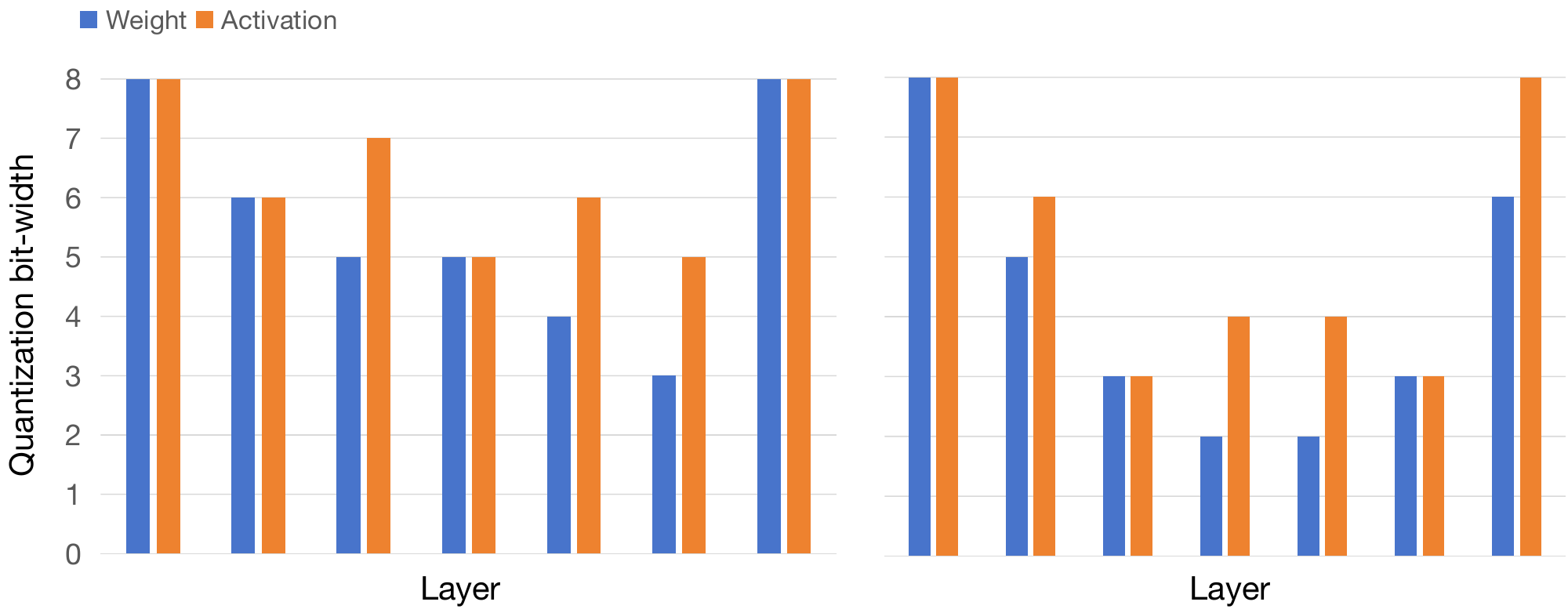}}
\caption{Quantization configuration searched by EdMIPs and SIMD-aware quantization explorer respectively}
\vspace{-1.5em}
\label{fig:quantization_search}
\end{figure}

In order to evaluate our hardware-aware quantization explorer, we choose EdMIPs as baseline, and utilize them to perform a search for optimal model quantization configurations. EdMIPs estimates complexity by using MACs as a proxy approximately. In contrast, according to Eq.\ref{eq:total_complexity}, our hardware-aware quantization explorer categorizes various operations within the operators, and adapt them with adjusting parameters. Quantization configurations searched by EdMIPs and our quantization explorer are illustrated in Fig. \ref{fig:quantization_search}. Compared to the quantization configuration searched by EdMIPs, our approach allows for quantizing to lower average bitwidths for both weights and activations under the same model architecture. Under the respective given quantization biwidth configurations, our model can reach up to 78.3\% Top-1 accuracy, which is +2.3\% up to EdMIPs, reflecting the effectiveness of our performance prediction model which can accurately directs NAS to perform hardware-aware quantization.

\section{Conclusion}
In this work, we present MCU-MixQ, a HW/SW co-optimized MPNN framework designed for MCU, which improves inference speed while meeting stringent hardware resources. We enhance the parallelism of the low-bitwidth convolution operator through packing and SIMD instructions, and meanwhile implement a low-bitwidth convolution library designed for MCU. As for model quantization, we employ differentiable NAS to automatically configure the optimal combination of quantization bit-widths for the model, while simultaneously considering the runtime efficiency of the model running on MCU. After quantization search stage, MPNN will undergo quantization-aware training and be ultimately mapped onto the optimized MCU kernels. Our experimental results demonstrate that MCU-MixQ achieves better performance compared to sota TinyEngine framework.
\par

\end{document}